\documentclass[%
 reprint,
preprintnumbers,
 amsmath,amssymb,
 aps,
prd,
]{revtex4-2}

\usepackage{graphicx}
\usepackage{dcolumn}
\usepackage{bm,xcolor,comment}

\def\E{\textit{\tiny{E}}} 
\def\H{\textit{\tiny{H}}} 
\def\HB{\textit{\tiny{H+dB}}} 
\def\nab{{\scriptscriptstyle{\nabla}}} 
\def\B{\textit{\tiny{B}}} 
\newcommand{\dd}{\mathrm{d}}

\newcommand{\gencom}[2]{[\![#1,#2]\!]}

\newcommand{\genpair}[2]{\eta(#1,#2)}
\newcommand{\genmet}[2]{\mathcal G( #1,#2)}
\newcommand{\gentor}[2]{T(#1,#2)}

\begin{document}
\preprint{Imperial-TP-2026-CH-2}
\title{Gravity and generalised geometry  from a Lie 2-algebroid perspective}

\author{Athanasios Chatzistavrakidis}
\email{athanasios.chatzistavrakidis@irb.hr}
\affiliation{Division of Theoretical Physics, Rudjer Bo\v skovi\'c Institute, Bijeni\v cka 54, 10000 Zagreb, Croatia}

\author{Chris Hull}
\email{c.hull@imperial.ac.uk}
\affiliation{Abdus Salam Centre for Theoretical Physics, Imperial College, London SW7 2AZ, United Kingdom}
\affiliation{London Institute for Mathematical Sciences, Royal Institution, London, W1S 4BS, United Kingdom}

\author{Larisa Jonke}
\email{larisa@irb.hr}
\affiliation{Division of Theoretical Physics, Rudjer Bo\v skovi\'c Institute, Bijeni\v cka 54, 10000 Zagreb, Croatia}
\affiliation{School of Theoretical Physics,  Dublin Institute for Advanced Studies,
 10 Burlington Road,  Dublin 4,  Ireland}

\author{Sylvain Lavau}
\email{slavau@gsu.edu.tr}
\affiliation{Department of Mathematics, Galatasaray \"Universitesi,  \c{C}\i ra\u{g}an Caddesi No:36, 34349 \.{I}stanbul, T\"urkiye}

\author{Peter Schupp}
\email{pschupp@constructor.university}
\affiliation{School of Science, Constructor University, Campus Ring 1, 28759 Bremen, Germany}

\begin{abstract}
We revisit the problem of defining natural torsion and curvature tensors for generalised connections and of using them to reconstruct the  effective action for the massless modes of the closed string from generalised geometry. The starting point is the description of a Courant algebroid as a differential graded (dg) symplectic manifold. A systematic procedure for split dg manifolds singles out a generalised Lie bracket as part of a larger Lie 2-algebroid structure, with the right properties to unambiguously define torsion and Riemann curvature tensors together with their corresponding Bianchi identities. This framework eliminates the structural circularity and ambiguity problems of previous approaches in the literature. We show that a unique generalised metric compatible generalised connection with fixed (con)torsion exists for any choice of ordinary connection that splits the dg manifold, while at the same time all Lie 2-algebroids obtained in this way are equivalent up to $L_{\infty}$ quasi-isomorphisms. For a given Lie 2-algebroid we demonstrate that the unique compatible generalised connection reproduces the effective closed string action from a geometric action based on the new generalised Ricci tensor upon making a canonical choice of torsion.

\end{abstract}

\maketitle


\section{Introduction}

Generalised geometry is a unifying extension of classical differential geometry that combines aspects of Riemannian, symplectic, and complex geometry into a framework that elegantly encodes both diffeomorphisms and higher gauge transformations as generalised diffeomorphisms acting on an extended bundle. The canonical example of such an extended bundle is the generalised tangent bundle $E = TM \oplus T^*M$, which formally combines vector fields and differential forms. There is a natural fiberwise pairing (metric) $\eta$ and an anchor map $\rho: E \rightarrow TM$ relating it to classical differential geometry and allowing the introduction of generalized Lie brackets and connections~\cite{courant1990dirac,liu1997manin,vsevera2017letters,hitchin2003generalized,gualtieri2011generalized}.  

Motivated in part by duality symmetries in string theory, generalized geometry has since become a natural setting for the description of flux backgrounds and the formulation of gravitational theories~\cite{Grana:2004bg}. Notably, it has been used in the context of supergravity, string and M-theory~\cite{Coimbra:2011nw,Hull:2007zu,Garcia-Fernandez:2013gja,Coimbra:2011ky}. Generalized geometry knows about certain stringy symmetries and greatly restricts the supergravity actions that can be formulated in this framework (see e.g.~\cite{Kupka:2024xur}), so that it is important for target space approaches to effective actions. 

There are, however, open problems with the appropriate definition of torsion and curvature tensors with nice properties and the determination of unique connections: There is, so far, no generalised analogue of the fundamental theorem of Riemannian geometry (although weaker versions exist, see e.g. \cite{Garcia-Fernandez:2013adv}). Often a pragmatic attitude is taken along the lines that there may be an infinite family of generalised connections satisfying certain metricity and torsion conditions, but most of them  lead to the same gravitational actions after applying suitable tweaks. Corrections are added to tensors as needed, projectors are introduced, etc. This is, of course, fine as long as one is primarily interested in an elegant way to construct actions, but questions arise if one wants to give a direct geometric interpretation to generalised curvature and torsion tensors. 

There are ways to avoid these ambiguities, and attempts have also been made to solve them: 
An elegant approach based on half-densities manages to bypass the need to define curvature tensors in the first place~\cite{Severa:2018pag}. In a Palatini type approach, unique connections follow from a variational principle, but the ambiguities are shifted to the definition of curvature and torsion~\cite{Jurco:2022nlr}. An approach based on deformations (by the fields of the theory) avoids the need to impose generalised metricity and torsion conditions and bypasses the accompanying ambiguities~\cite{Boffo:2019zus,Boffo:2021srg}.

In the current work we address these problems, show how to determine unique (up to a choice of ordinary connection) generalised connections, define curvature and torsion tensors with properties similar to the ones found in Riemannian geometry, and use them to construct gravity actions without the need to correct tensors or introduce projectors. The extra ingredient is the structure of a Lie 2-algebroid that nicely complements the structure of the Courant algebroids of ordinary generalised geometry. This Lie 2-algebroid structure emerges from the choice of a $TM$ connection on $E$, splitting the graded manifold $T^*[2]E[1]$ into a (more amenable) graded vector bundle~{\footnote{In the literature, Lie 2-algebroids are often identified with dg manifolds of degree 2, not necessarily split, and \emph{split} Lie 2-algebroids correspond to split dg manifolds. We do not follow strictly this nomenclature and what we call a \emph{Lie 2-algebroid} here is a \emph{split} one in the sense of~\cite{Sheng_2017}.}}.

An alternative approach to capturing stringy symmetries is double field theory, which formally doubles spacetime coordinates to make dualities manifest~\cite{Hull:2009mi}. Several constructions of generalised geometry have been motivated by related constructions in double geometry \cite{Hohm:2010pp,Hohm:2011si,Hohm:2012mf,Polacek:2013nla,Hassler:2024yis,Hassler:2025rag}. Here we refrain from doubling the base manifold and defer such considerations to future work.

\subsection{Two brackets, torsion and curvature}

The Lie bracket of vector fields 
$\mathcal L_X Y \equiv [X,Y]_\text{Lie}$
enjoys dual roles in classical differential geometry, arising  as both the Lie derivative  of a vector field along another vector field and as the commutator of vector fields.
Seeking analogues of the Lie bracket in generalised geometry gives rise to two distinct brackets, one of which is a generalised Lie derivative while the other is a generalisation of the commutator.
 We shall discuss both in detail in this paper. 
 
 The role of Lie derivative in generalised geometry is taken by the Dorfman bracket, i.e.\ a bracket satisfying the axioms of a Courant algebroid (see section \ref{CAsection}) that we will denote by $\mathcal L_e e' \equiv [e,e']$ for $e,e' \in \Gamma(E)$: It satisfies the Jacobi identity and the Leibniz rule in the second slot,
it is  compatible with the fibre metric $\eta$, and its $H$-twisted version is covariant under $O(d,d)$ transformations -- i.e.\ it has all the right properties. Unlike the Lie bracket, it is not skew symmetric and does not satisfy the Jacobi identity and the Leibniz rule in the first slot:
\begin{align}
[e,e] &= D \genpair{e}{e}\,, \label{CAaxiom3}\\
[f e , e'] &= - \rho(e') f \, e + f\, [e, e'] + \genpair{e}{e'}Df \,,
\label{modLeib}
\end{align}
 where $Df$ is defined via $\genpair{Df}{e} = \rho(e) f$. The extra terms involving $\eta$ are zero in classical differential geometry as the pairing $\genpair{X}{Y}$ 
of two vector fields is canonically zero, but they turn out to be essential in the following.

Let us now focus on the second incarnation of the Lie bracket: we introduce a \lq generalised Lie bracket' $\gencom{e}{e'} \in \Gamma(E)$  that will appear in the expressions for generalised torsion and curvature tensors of an $E$-connection~$\nabla^{\E}$ on~$E$~\cite{Boffo:2019zus}: 
\begin{align}
T(e,e') &= \nabla^\E_e e' - \nabla^\E_{e'} e - \gencom{e}{e'}\,, \\
R(e,e') &= \nabla^\E_e \nabla^\E_{e'} - \nabla^\E_{e'} \nabla^\E_e -
\nabla^\E_{\gencom{e}{e'}}\,.
\end{align}
Tensoriality of $T$ and $R$ requires that the new bracket must satisfy Leibniz rules in both slots, 
\begin{align}
\gencom{e}{f e'} &= \rho(e) f \, e' + f \gencom{e}{e'} \,,\\
\gencom{f e}{e'} &= -\rho(e') f \, e + f \gencom{e}{e'} \,,
\end{align}
which is clearly not the case for the Dorfman bracket or for its antisymmetrized version, the Courant bracket.

We point out that given a generalised Lie bracket and a generalised and not necessarily symmetric metric $\genmet{e}{e'}$, we can derive an analogue of the fundamental theorem of Riemannian geometry with the generalised  Koszul formula
\begin{equation}
\begin{split}
&2 \mathcal G_{\text{sym}}(e'',\nabla^\E_e e') 
\equiv \genmet{e''}{\nabla^\E_e e'} + \genmet{\nabla^\E_e e'}{e''} \\
& \quad  = \rho(e')\genmet{e''}{e} + \rho(e)\genmet{e'}{e''} - \rho(e'')\genmet{e'}{e} \\
& \quad  + \genmet{e''}{\gencom{e}{e'}}- \genmet{e'}{\gencom{e}{e''}} - \genmet{\gencom{e'}{e''}}{e}\\
& \quad + 2 K(e'',e,e') + 2N(e'',e, e')\, .
\label{genkoszul}
\end{split}
\end{equation}
Here the contorsion and distortion (non-metricity) tensors are
\begin{align}
K(&e'',e,e') =\nonumber\\ &\tfrac12\big(\genmet{e''}{\gentor{e}{e'}}- \genmet{e'}{\gentor{e}{e''}} - \genmet{\gentor{e'}{e''}}{e}\big)\,, \\
N(&e'',e,e') =\nonumber\\ &\tfrac12\big(\nabla^\E_{e''}\genmet{e'}{e}-\nabla^\E_{e'}\genmet{e''}{e} - \nabla^\E_{e}\genmet{e'}{e''} \big)\,.
\end{align}
Choosing tensors $K$ and $N$ and assuming non-degeneracy of $\mathcal G_{\text{sym}}$, this formula gives a unique expression for the generalised connection $\nabla^\E$ in terms of $\mathcal G$ and its derivatives. Clearly, there must not be an explicit dependence of the generalised Lie bracket on the connection $\nabla^\E$ for this to work.

A main result of the current paper is the identification of a natural candidate for the generalised Lie bracket that is admissible in this sense, avoiding the circular definitions of some previous approaches. We will not need \eqref{genkoszul} in full generality, since the usual generalised metric~$\mathcal H$, while being based on the non-symmetric combination $g+B$, is in fact symmetric and hence $\mathcal H  = \mathcal G=  \mathcal G_{\text{sym}}$.

Comparing the two generalizations of the Lie brackets, we find that their difference defines an $E$ on $E$ connection~$\dot\nabla^\E$ (that has a priori no relationship to the one that appears in the Koszul formula) \cite{Boffo:2019zus}:
\begin{equation}
\genpair{e}{[e',e'']-\gencom{e'}{e''}} = \genpair{\dot\nabla^\E_e e'}{e''}\,.
\label{tripleid}
\end{equation}
$C^\infty$-linearity in the first and third slot of (\ref{tripleid}) ($e$ and~$e''$) follows from the $C^\infty$-bilinearity of the fibre metric $\eta$ and the Leibniz property of both brackets in their second slots. The expression is thus well defined and satisfies one of the properties of a generalised connection. The second property,
$\dot\nabla^\E (f e') = df \, e' + f\, \dot\nabla^\E e'$,
follows from \eqref{modLeib}. 

The symmetric part of $\gencom{-}{-}$ is tensorial 
and we can thus consistently set it equal to zero. Via \eqref{CAaxiom3}, skew  symmetry of $\gencom{-}{-}$ is equivalent to $\dot\nabla^\E$ being compatible with the fibre metric $\eta$ and we shall henceforth assume this. The remaining freedom in the choice of $\gencom{-}{-}$ is an antisymmetric 2-tensor that can be absorbed into the choice of contorsion tensor. A skew symmetric bracket with Leibniz rule has also been studied in~\cite{lean2018dorfman,jotz_modules_2023}, but without the further structures that we require in the present work.

For a given choice of generalised contorsion and distortion we thus find a unique generalised connection, \emph{provided that $\gencom{-}{-}$ does not explicitly depend on $\nabla^\E$}. But this requirement is exactly the problem with existing definitions of generalised torsion~\cite{10.1093/acprof:oso/9780199534920.003.0018} and curvature: They effectively define $\gencom{-}{-}$ in terms of $[-,-]$ by setting $\dot\nabla^\E = \nabla^\E$ in  \eqref{tripleid} with the result that \eqref{genkoszul} has no longer a unique solution (if it has a solution at all.)

In this paper we take a different route and argue that the natural choice for the generalised Lie bracket is the skew symmetric bracket $\gencom{e}{e'} =\ell_2(e,e')$ of a \emph{Lie 2-algebroid}, which satisfies a Jacobi identity up to homotopy and leads to Bianchi identities for the generalised torsion and curvature tensors that resemble those of classical differential geometry. 
From the perspective of graded geometry, the Dorfman bracket $[e,e']$ can be understood as a derived bracket. In the same setting, there is another natural construction which gives $\ell_2(e,e')$ (as well as $\ell_1, \ell_3$) and gives an explanation for \eqref{tripleid} with $\dot\nabla^\E$ arising as a derived connection. For the sake of completeness, by Lie 2-algebroid, we designate here a split graded vector bundle concentrated in degrees $0$ and $-1$, equipped with a Lie 2-algebra structure on its sheaf of sections compatible with the vector bundle structure. We refer to \cite{ladaIntroductionSHLie1993, ladaStronglyHomotopyLie1995} for the original references on $L_\infty$-algebras, and to \cite{lavauetal} for a description of Lie $n$-algebroids in the symmetric (shifted) convention. Section \ref{sec:ref} clarifies how to pass from the symmetric (shifted) to the skew symmetric (unshifted) picture.

\section{Courant \& Lie 2-algebroids} 

\subsection{Courant algebroids \& generalised connections} 
\label{CAsection}

We begin with a brief description of the basic notions that constitute the backbone of our construction. A Courant algebroid consists of a vector bundle $E$ on a smooth manifold $M_0$ together with (i) a binary operation, which we denote as a bracket $[-,-]: \Gamma(E)\times \Gamma(E)\to \Gamma(E)$ on the vector space of sections $\Gamma(E)$, not necessarily skew symmetric, (ii) a nondegenerate symmetric bilinear form (a metric on $E$) $\eta: \Gamma(E)\times \Gamma(E)\to C^{\infty}(M_0)$ and (iii) a smooth map $\rho: E\to TM_0$. We refer to these three structures as the Dorfman bracket, the inner (or background) metric and the anchor map, respectively. They are such that the pair $(\Gamma(E),[-,-])$ is a (left) Leibniz algebra, which means that for $e, e', e''\in \Gamma(E)$ we have the Leibniz identity
\begin{equation}
[e,[e', e'']]=[[e, e'], e''] +[e', [e, e'']]\,.    
\end{equation} 
In addition, two more properties complete the definition:  
\begin{equation}
\eta(e,[e',e'])=\tfrac 12 \rho(e)\eta(e',e')=\eta([e, e'],e')\,.
\end{equation}
These three equations imply the following two \cite{Uchino}:
\begin{align}
    [e,fe']&=f[e,e']+\rho(e)f\, e'\,,
    \label{Leibniz for functions}
    \\[4pt]
    \rho([e, e'])&=[\rho(e),\rho(e')]_{\text{Lie}}\,,\label{eq:homomorphismproperty}
\end{align}
for any $f\in C^{\infty}(M_0)$, where on the right hand side of~\eqref{eq:homomorphismproperty} the bracket is the usual Lie bracket of vector fields. Note that since the bracket has a symmetric part, the right Leibniz rule for functions \eqref{Leibniz for functions} does not imply that a left Leibniz rule for $[fe, e']$ holds. Indeed the result in this case contains additional terms that involve the metric $\eta(e,e')$ (see \eqref{modLeib}), and the same holds for the antisymmetrised version of the bracket (called the Courant bracket), which contains such terms in both the left and the right Leibniz rules for functions.   

As for any vector bundle, it is possible to define a connection (covariant derivative) on a Courant algebroid as a map $\nabla: \mathfrak{X}(M_0)\times \Gamma(E)\to \Gamma(E)$, with $\mathfrak{X}(M_0)$ denoting the sheaf of vector fields on the manifold. We will call such maps \emph{$TM_0$-on-$E$ connections} or \emph{ordinary connections}. What is more, it is possible to define a notion of \emph{$E$-on-$E$} or \emph{generalised connection} with the help of the anchor. This is a map $\nabla^{\E}: \Gamma(E)\times \Gamma(E)\to \Gamma(E)$, which is homogeneous, linear in the first slot and satisfies the Leibniz rule for $f\in  C^{\infty}(M_0)$: 
\begin{align}    \nabla^{\E}_{fe}e'&=f\,\nabla^{\E}_{e}e'\,,  \\[4pt] \nabla^{\E}_{e}(fe')&=f\,\nabla^{\E}_{e}e' +\rho(e)f\, e'\,. 
\end{align}
If, in addition, a generalised connection satisfies $\nabla^{\E}\eta=0$, we call it \emph{$\eta$-metric} or \emph{$\eta$-compatible}. We can also have $\eta$-metricity for an ordinary connection when $\nabla\eta=0$. 

It is useful to express these connections in a local coordinate system. Denote coordinates on $M_0$ as $x^{\mu}$ and a local frame of $E$ as $\mathfrak{e}_a$, with $a=1,\dots,\text{rk}\,E$. We denote the components of $[-,-], \eta, \rho$ as $C_{ab}{}^{c}, \eta_{ab}, \rho_{a}{}^{\mu}$, respectively. We may then express the connections $\nabla$ and $\nabla^{\E}$ as: 
\begin{align}
\nabla_{\mu}e^{b}&=\partial_{\mu}e^{b}+\Gamma_{\mu a}{}^{b}e^{a}\,, \\[4pt]
    \nabla^{\E}_{a}e^{b}&=\rho_{a}{}^{\mu}\partial_{\mu}e^{b}+\omega_{ac}{}^{b}e^{c}\,.
    \end{align}
Then $\eta$-metricity for the generalised connection amounts to the antisymmetry of its coefficients in the last two indices,  
\begin{equation}
    \omega_{abc}=-\,\omega_{acb}\,, \label{eta metricity omega}
\end{equation}
where $\omega_{abc}=\eta_{cd}\omega_{ab}{}^{d}$, and similarly for the ordinary connection, 
\begin{equation}
    \Gamma_{\mu ab}=-\Gamma_{\mu ba}\,, 
\end{equation}
once again lowering Latin indices with the metric $\eta$.

\subsection{Courant algebroids as QP2 manifolds}

An elegant reformulation of Courant algebroids uses the language of graded manifolds \cite{Roytenberg:1999mny,Roytenberg:2002nu}. A Courant algebroid is described as a differential graded (dg) symplectic manifold of degree $2$ (sometimes called QP2 manifold), a higher analogue of the more familiar Poisson structures which correspond to degree $1$ graded symplectic manifolds. In this sense, Courant algebroids are to strings what Poisson structures are to particles. 

This approach introduces two sets of coordinates and conjugate momenta, which we will denote as $(x^{\mu},p_{\mu})$ and $(\psi^{a},\xi_{a})$, respectively. Note that the first pair corresponds to coordinates on $M_0$ and conjugate momenta for them, suggesting a phase space structure, in other words a cotangent bundle. The second pair corresponds to vector bundle coordinates on $E$, as it is clear from their Latin index, and their own conjugate momenta. We may now merge these pairs using the structure of a graded manifold, by first assigning degree $1$ to $\psi^{a}$, which are therefore fermionic, anticommuting coordinates. They are now fiberwise coordinates on the parity-shifted vector bundle $\Pi E \equiv E[1]$, whose coordinates as a manifold are thus $(x^{\mu},\psi^{a})$, and whose sections are consequently considered to be of degree $-1$. To include the conjugate momenta, we take the phase space $T^*E[1]$, i.e. the cotangent bundle of $E[1]$, to which we impose an additional degree shift by $-2$. This results in local coordinates $(x^{\mu},\psi^{a},\xi_{a},p_{\mu})$ of respective degrees $(0,1,1,2)$ on the corresponding graded manifold $T^{\ast}[2]E[1]$.    
Then the QP2 manifold $M_2$ associated to the Courant algebroid~$E$ is defined as the set of points of $T^\ast[2]E[1]$ satisfying $\xi_a=\tfrac{1}{2}\eta_{ab}\psi^b$, and it is called the minimal symplectic realisation of $E[1]$. More abstractly, it is the pull-back of $T^{\ast}[2]E[1]$ over the map $E\to (E\oplus E^*)[1], e\mapsto e\oplus \tfrac{1}{2}\eta(e,\cdot\,)$ \cite{Roytenberg:2002nu}. For the case $E=TM_0\oplus T^*M_0$, we find that $M_2$ is isomorphic to $T^*[2]T[1]M_0$, and more generally, for a Courant algebroid~$E$ we have that $M_2$ is (non-canonically) isomorphic to $T^{\ast}[2]M_0\oplus E[1]$.

As a (graded) cotangent bundle, $T^*[2]E[1]$ is canonically endowed with a symplectic form of degree 2, 
\begin{equation}
    \Omega=dp_\mu\wedge dx^\mu+\tfrac{1}{2}\eta_{ab}\,d\psi^a\wedge d\psi^b\,,
\end{equation}
which restricts to a symplectic form on $M_2$.
This gives rise to the P(oisson)-structure, with Poisson brackets~$ \{ .,.\}$ of degree $-2$. In addition, there is a Q-structure, a vector field of degree 1 that satisfies $[Q,Q]=0$ with respect to the graded commutator of vector fields (it is homological). In the above coordinate system, the compatibility of the Q and P structures -- which is encoded by the identity $\mathcal{L}_Q(\Omega)=0$ -- restricts the homological vector field to the form 
\begin{eqnarray}
Q&=&-\eta^{ab}\rho_a{}^\mu p_\mu\frac{\partial}{\partial \psi^b}+\rho_a{}^\mu \psi^a\frac{\partial}{\partial x^\mu}-\tfrac{1}{2}\eta^{ad}C_{bcd}\psi^b\psi^c\frac{\partial}{\partial \psi^a} \nonumber\\
&&-\,\partial_\mu\rho_a{}^\nu \psi^a p_\nu\frac{\partial}{\partial p_\mu}-\tfrac{1}{6}\partial_\mu C_{abc}\psi^a\psi^b\psi^c\frac{\partial}{\partial p_\mu}\,, \label{Q}
\end{eqnarray}
where the components of the Dorfman bracket, inner metric and anchor map are recognised. This allows us to define a degree 3 Hamiltonian function 
\begin{equation}
    \Theta=-\,\psi^a\rho_a{}^\mu p_\mu-\tfrac{1}{6}C_{abc}\,\psi^a\psi^b\psi^c\,, 
    \label{Hamiltonian QP2}
\end{equation}
which satisfies the Maurer-Cartan equation $\{\Theta,\Theta\}=0$. It is related to the homological vector field   by:
\begin{equation}\label{eq:homologicalvector}
    Q=\{\Theta,-\}\,.
\end{equation}
Using the Poisson bracket and the Hamiltonian, the data of a Courant algebroid are elegantly encoded as 
\begin{subequations}
\begin{eqnarray}
    [e,e']&=&\{\{\Theta,e\},e'\}\,, 
    \\[4pt] \rho(e)f&=&\{\{\Theta,e\},f\}\,, \\[4pt] \eta(e,e')&=&\{e,e'\}\,,
\end{eqnarray}
\end{subequations}
by means of the identification of degree 1 functions and sections $C_{1}(M_2)=\Gamma(E^{\ast})\simeq\Gamma(E)$.   

\subsection{The Lie 2-algebroid construction}\label{sec:ref}

We now motivate an alternative construction, foreshadowing our approach in the next section. Note that since the Dorfman bracket does not satisfy the Leibniz rule for functions in its first entry, and furthermore is not skew symmetric, it is not a suitable basis for a definition of a torsion tensor for generalised connections in a natural way. Skew symmetrizing it, the Courant bracket is obtained, which is also not suitable for this task because it still does not satisfy the usual Leibniz rule for functions in either of its entries. One may ask: \emph{is there any bracket that satisfies both properties of skew symmetry and Leibniz rule in both entries for functions?} 

To answer this question and identify such a bracket, we look at an alternative approach to describing Courant algebroids. It is generally known that for a differential graded manifold of degree $n$ (not necessarily symplectic), it is possible to construct a structure called a Lie $n$-algebroid.
A Lie $n$-algebroid is to a Lie $n$-algebra what a Lie algebroid is to a Lie algebra---the same structure on the space of sections together with an anchor map. In that case, a Lie $n$-algebroid is a chain complex of (graded) vector bundles, 
\begin{equation}\label{eq:degree}
E_{-n+1}\longrightarrow E_{-n+2}\longrightarrow\ldots\longrightarrow E_{-1}\longrightarrow E_0\,,
\end{equation}
each of them carrying a degree between $-n+1$ and $0$, and equipped with a family of multilinear brackets satisfying higher versions of the Jacobi identity.
This convention ensures that for a Lie 1-algebroid, i.e.\ a Lie algebroid, the only vector bundle appearing is $E_0$, concentrated in degree 0.

The construction relating the dg-manifold structure to the Lie $n$-manifold structure only works for \emph{split} (non-negatively graded) dg manifolds, i.e. precisely for those which consist of families of (non-positively graded) vector bundles over $M_0$ \cite{Bonavolont__2013}, namely \emph{graded vector bundles}.
However, a Courant algebroid seen as the QP2 manifold~$M_2$ is \emph{not split}, in general. More precisely, although the degree 1 coordinates $\psi^a$ and $\xi_a$ on $M_2$ transform as tensors, the degree 2 local coordinates $p_\mu$ do not. Indeed, they transform affinely under a change of coordinates on the base manifold $M_0$, as their transformation law involves a contribution from the product of $\psi^a$ and $\xi_a$ \cite{Roytenberg:2002nu}.

In order to linearize this transformation law we can use a section of the projection $M_2\longrightarrow E[1]$. There is no canonical section, but any choice results locally in the structure of a $TM_0$-on-$E$ connection that is $\eta$-metric. The connection coefficients are then used 
to define transformed momenta $\widetilde{p}_\mu=p_\mu+\Gamma_{\mu a}{}^b\psi^a\xi_b$, which then behave as tensors. $\eta$-metricity is required to ensure the covariance of the equation $\xi_a=\tfrac{1}{2}\eta_{ab}\psi^b$ defining the submanifold $M_2$ inside $T^*[2]E[1]$. As a result, this choice of connection has the effect of `splitting' the graded manifold $M_2$ into the Whitney sum $T^*[2]M_0\oplus E[1]$, a graded vector bundle over $M_0$. 

Alternatively, let us provide a more mathematical and non-local justification of the above argument. Upon a choice of $\eta$-metric $TM_0$-on-$E$ connection, we observe that the sheaf of degree 2 functions on $M_2$ is non-canonically isomorphic to the sheaf of sections of the vector bundle $\wedge^2E^*\oplus TM_0$. On the other hand, the latter 
sits inside an exact sequence of vector bundles, called the \emph{Atiyah sequence}:
\begin{equation}\label{eq:exactatiyah}
0\longrightarrow \mathfrak{so}(E)\simeq\wedge^2 E^*\longrightarrow \mathbb{A}\longrightarrow  TM_0\longrightarrow 0\,.
\end{equation}
The vector bundle $\mathbb{A}$ is a Lie algebroid over $M_0$ whose sections consist in the covariant differential operators on~$E$ preserving the metric $\eta$~\cite{Kosmann2002, Roytenberg:2002nu}. The sheaf of sections of~$\mathbb A$ is then isomorphic to the sheaf
of degree 2 functions on~$M_2$, and a splitting of the exact sequence \eqref{eq:exactatiyah} is precisely an $\eta$-metric $TM_0$-on-$E$ connection. 

Suspending the degrees of the fibers of the graded vector bundle $T^{\ast}[2]M_0\oplus E[1]$ by $+1$ (in order to be consistent with the Lie $n$-algebroid setup \eqref{eq:degree}), we are able to construct a Lie 2-algebroid structure on $T^{\ast}[1]M_0\oplus E$. 
The construction begins with the unweighted Euler vector field $\varepsilon$ on $M_2$, which counts the number of graded generators (the \emph{homological degree}). This is not a global vector field; in the coordinate system we introduced, it takes the form 
\begin{equation}
    \varepsilon=\psi^a\frac{\partial}{\partial \psi^a}+p_\mu\frac{\partial}{\partial p_\mu}\,,
\end{equation}
and it is not respected by general coordinate transformations. However, a global unweighted Euler vector field exists in the split case. Its explicit form is 
\begin{align}
\varepsilon^{\scriptscriptstyle{\nabla}}=\psi^{a}\big(\frac{\partial}{\partial\psi^{a}}+\Gamma_{\mu ba}\psi^{b}\frac{\partial}{\partial p_{\mu}}\big)+\big(p_{\mu}+\tfrac 12 \Gamma_{\mu ab}\psi^{a}\psi^{b}\big)\frac{\partial}{\partial p_{\mu}}\,, \nonumber \\[4pt]
\end{align}
where we have denoted its dependence on the connection with a superscript $^{\scriptscriptstyle\nabla}$.
We would now like to compute the graded commutator $[\varepsilon^{\scriptscriptstyle\nabla},Q]$, where $Q$ is the homological vector field on $M_2$. 
A direct calculation establishes that we can write 
\begin{equation}
[\varepsilon^{\scriptscriptstyle\nabla},Q]=\sum_{s=0}^{2}s\, Q^{(s)}\,,
\end{equation}
where $s$ is the homological degree and $Q=Q^{(0)}+Q^{(1)}+Q^{(2)}$. 
This decomposes $Q$ into terms with $s$ graded generators. In this decomposition with respect to the global Euler vector field, each eigenvector depends on the connection. Specifically, $Q^{(0)}$ corresponds to the first term in \eqref{Q}, $Q^{(2)}$ to the last term and $Q^{(1)}$ to the remaining three terms, all covariantised using the connection $\nabla$.

After decomposing the homological vector field into homological degrees, the following set of (higher derived) graded symmetric brackets \cite{Voronov2005jppa}
\begin{equation}\label{b brackets}
    b_{s+1}(e_1,\dots,e_{s+1})=[\dots[[Q^{(s)},e_1],e_2]\dots e_{s+1}]\,,
\end{equation}
 gives rise to a (shifted) Lie 2-algebroid structure on $T^\ast[2]M_0\oplus E[1]$ with all brackets of the same degree, namely $1$ (this is called the ``$b$-picture'' in~\cite{Hohm:2017pnh}). By applying to the present case -- i.e. for $s=0,1,2$ -- the canonical formula \cite{FiorenzaManetti} relating degree~$1$ symmetric brackets on $T^\ast[2]M_0\oplus E[1]$ to  degree~$2-s$ skew symmetric brackets on $T^\ast[1]M_0\oplus E$ (the ``$\ell$-picture'' in \cite{Hohm:2017pnh}, up to their convention that their degree is $s-2$), we obtain a Lie 2-algebroid structure with graded skew symmetric brackets on $T^{\ast}[1]M_0\oplus E$. In other words, a set of graded skew symmetric brackets $\ell_1, \ell_2, \ell_3$ of arity 1, 2, 3 and degree $1, 0, -1$ respectively, such that $\ell_1$ and $\ell_3$ are $C^{\infty}(M_0)$-linear and $\ell_2$ is skew symmetric. As a consequence of the first term in $Q^{(1)}$, $\ell_2$ satisfies the Leibniz rule in both entries 
 \begin{subequations}
\begin{align}\label{eq:Leibnizpropertyl2}
    \ell_2(e,f e')&=f\ell_2(e,e')+\rho(e)f\, e'\,,
    \\[4pt] 
    \ell_2(fe, e')&=f\ell_2(e,e')-\rho(e')f \,e\,,
\end{align}
\end{subequations}
with respect to the anchor $\rho: E\to TM_0$. Moreover they satisfy the identities 
\begin{subequations}
\begin{align}\label{Linf relations 1}
   & \ell_1\ell_2=\ell_2\ell_1\,, \\[4pt] &\ell_2\ell_2+\ell_1\ell_3+\ell_3\ell_1=0\,, \label{Linf relations 2}
   \\[4pt] &\ell_2\ell_3=\ell_3\ell_2\,, \label{Linf relations 3}
\end{align}
\end{subequations}
where in both terms $\ell_2\ell_1$ and $\ell_3\ell_1$, the unary bracket $\ell_1$ is applied successively on each argument, with the property that $\ell_1(E)=0$ and that $\ell_3$ is non-zero only on $\Gamma(\wedge^3E)$.
Notice that \eqref{Linf relations 2}, together with Equation~\eqref{eq:Leibnizpropertyl2}, imply the  identity
\begin{equation}
    \rho(\ell_2(e,e'))=[\rho(e),\rho(e')]_{\text{Lie}}\,,
\end{equation}
which is reminiscent of Eq. \eqref{eq:homomorphismproperty}.

Accordingly, we have a chain complex concentrated in only two degrees $-1$ and $0$: 
\begin{equation}\label{eq:degree special case}
 E_{-1}=T^{\ast}[1]M_0\longrightarrow E_0=E\,.
\end{equation}
Since $\ell_1$ has degree $1$, it only acts nontrivially on 1-forms (degree $-1$) and produces an $E$-section (degree $0$). Similarly, since $\ell_2$ has degree $0$ and takes two entries, it has two nontrivial components: the first is to take two $E$-sections and produce an $E$-section, and the second is to take one $E$-section and one 1-form and produce a $1$-form. Finally, since the $\ell_3$ bracket has degree $-1$, it can only take three $E$-sections and produce a $1$-form. These saturate the nontrivial possibilities.

The various brackets are computed using the eigenvectors $Q^{(s)}$ of the global unweighted vector field in \eqref{b brackets} and switching to the $\ell$-picture. 
First we report the final result in coordinate-independent form:
\begin{subequations}
\begin{eqnarray}
\ell_1(\lambda)&=&-\,\rho^\sharp(\lambda)\,, \\[4pt]
e^\ast\left(\ell_2(e,e')\right)&=&e^\ast\left([e, e']\right)- \eta(\dot\nabla^{\E}_{\eta^{\sharp}(e^{\ast})}e,e')\,, \label{ell 2}\\[4pt]
\ell_2(e,\lambda)(X)&=&L_{\rho(e)}\lambda (X)- \eta(\nabla_{X} e,\rho^\sharp(\lambda))\,,\\[4pt] 
    \ell_3(e,e',e'')(X)&=& -\,\eta(S^\nabla(e,e')(X),e'')\,,
\end{eqnarray}
\end{subequations}
where $\lambda\in \Gamma(T^{\ast}[1]M_0)$ is a $1$-form, $e^{\ast}\in\Gamma(E^{\ast})$ is a section of the dual vector bundle and $X\in\mathfrak{X}(M_0)$ is a vector field. Here, $\rho^{\ast}: T^{\ast}M_0\to E^{\ast}$ is the map dual to the anchor, $\rho^{\sharp}: T^{\ast}M_0 \to E$ is its composition with the isomorphism $\eta^\sharp:E^{\ast}\to E$ induced by the fiberwise metric~$\eta$, and $S^{\nabla}: \Gamma(E)\times \Gamma(E)\times \mathfrak{X}(M_0)\to \Gamma(E)$ is a tensor map, called the \emph{basic curvature} of the ordinary connection $\nabla$, which measures the compatibility of the $TM_0$-on-$E$ connection with the Dorfman bracket---we refer to~\cite{Chatzistavrakidis:2023otk} for its coordinate-independent definition and properties, since we do not use it further in the present work. We report, however, its form in local coordinates below. Finally, $\dot{\nabla}^{\E}$ is the canonical $E$-on-$E$ connection induced by~$\nabla$ through the anchor, specifically 
\begin{equation}\label{nabla dot}
    \dot{\nabla}^{\E}_ee':=\nabla_{\rho(e)}e'\,,
\end{equation} 
whose coefficients are $\dot{\omega}_{ab}{}^{c}=\rho_{a}{}^{\mu}\Gamma_{\mu b}{}^{c}$. This construction was previously reported in Ref.\,\cite{jotz_modules_2023} too, with different motivations.

It is also useful to write down the expressions of these brackets in local coordinates. They are
\begin{subequations}
\begin{align}
\ell_1(\lambda)^{a}=&-\eta^{ab}\rho_{b}{}^{\mu}\lambda_{\mu}\,, \\[4pt] 
\ell_{2}(e,e')^{a}=& \,e^b\rho_b{}^\mu\partial_\mu e'^a-e'^b\rho_b{}^\mu\partial_\mu e^a\nonumber\\[4pt]
&+C_{bc}{}^ae^be'^c-\eta^{ab}\rho_{b}{}^\mu\Gamma_{\mu cd}e^ce'^d\,,
\\[4pt] 
\ell_2(e,\lambda)_{\mu}=&\,e^a(\rho_a{}^\nu\partial_\nu\lambda_\mu+\lambda_\nu\partial_\mu\rho_a{}^\nu-\rho_b{}^\nu\Gamma_{\mu a}{}^b\lambda_\nu)\nonumber\\[4pt]
=&-\ell_2(\lambda,e)_{\mu}\,,
\\[4pt] 
\ell_{3}(e,e',e'')_{\mu}=&-e^{a}e'^be''^{c}S^{\scriptscriptstyle{\nabla}}_{abc\mu}\,,
\end{align}
\end{subequations}
where 
\begin{align}
S^{\scriptscriptstyle{\nabla}}_{abc\mu}=&\,\partial_\mu C_{abc}-3C_{d[bc}\Gamma_{\mu a]}{}^d 
-3\partial_\mu\rho_{[a}{}^\nu\Gamma_{\nu bc]} \nonumber\\[4pt] &-3\rho_{[a}{}^\nu\partial_\nu\Gamma_{\mu bc]} +3\rho_d{}^\nu\Gamma_{\nu [bc}\Gamma_{\mu a]}{}^d
\end{align}
are the components of the basic curvature for the ordinary connection on the Courant algebroid \cite{Chatzistavrakidis:2023otk}. 

Some important remarks follow: First and foremost, for the construction to result in a skew symmetric bracket~$\ell_2$, the connection $\nabla$ must be $\eta$-metric, which we always assume. Any other choice of connection $\nabla'$ that splits the QP2 manifold results in an $L_{\infty}$ quasi-isomorphic Lie 2-algebroid. It is also worth mentioning that there exists a canonical $L_{\infty}$ morphism between the Roytenberg-Weinstein Lie 2-algebra \cite{roytenbergCourantAlgebroidsStrongly1998}, which can be constructed for any Courant algebroid, and the Lie 2-algebra $\Omega^1(M)[1]\oplus \Gamma(E)$. We leave a detailed proof of these statements for a more mathematically oriented future work.

The upshot of this analysis is that the described construction results in a binary bracket $\ell_2(e,e')$, which is skew symmetric and satisfies the Leibniz rule for functions in both entries. This is true for any $\eta$-compatible connection $\nabla$, including the trivial connection. These are precisely the desired properties to define natural tensorial quantities for generalised connections.

\section{Torsion and curvature revisited}

\subsection{Torsion \& curvature from the Lie 2-algebroid}

The construction presented in the previous section makes the definition of torsion and curvature tensors for generalised connections completely straightforward: 
\begin{eqnarray}
\label{torsion} T_{\scriptscriptstyle\nabla}^{\scriptscriptstyle\nabla^{\E}}(e,e')&=&\nabla^{\E}_{e}e'-\nabla^{\E}_{e'}e-\ell_2(e,e')\,, \\[4pt]
\label{curvature} R_{\scriptscriptstyle\nabla}^{\scriptscriptstyle\nabla^{\E}}(e,e')&=& [\nabla^{\E}_{e},\nabla^{\E}_{e'}]-\nabla^{\E}_{\ell_2(e,e')}\,.
\end{eqnarray}
We have indicated with a subscript that these are given with respect to the choice of connection $\nabla$, since $\ell_2$ depends on it. That the above objects indeed give rise to tensors requires no proof; it is guaranteed by construction due to the properties of $\ell_2$. 

The naturalness of these definitions is further justified by the fact that we obtain the first and second Bianchi identities at face value, essentially due to the defining condition of the Lie 2-algebroid. They follow from the $L_{\infty}$ relations \eqref{Linf relations 1} and \eqref{Linf relations 2} and they read: 
\begin{eqnarray}
&& R_{\scriptscriptstyle\nabla}^{\scriptscriptstyle\nabla^{\E}}(e,e')e'' +R_{\scriptscriptstyle\nabla}^{\scriptscriptstyle\nabla^{\E}}(e'',e)e'+R_{\scriptscriptstyle\nabla}^{\scriptscriptstyle\nabla^{\E}}(e',e'')e= \nonumber\\ && \hspace{2cm}=\big(\dd^{\scriptscriptstyle\nabla^{\E}}T_{\scriptscriptstyle\nabla}^{\scriptscriptstyle\nabla^{\E}}+\ell_1\ell_3\big)(e,e',e'')\,,\label{BI1}\\[4pt]
&&    d^{\scriptscriptstyle\nabla^{\E}}R_{\scriptscriptstyle\nabla}^{\scriptscriptstyle\nabla^{\E}}+\nabla^{\E}_{\ell_1\ell_3}=0\,, \label{BI2}
\end{eqnarray}
where $\dd^{\scriptscriptstyle\nabla^{\E}}$ is the exterior covariant differential with respect to the generalised connection. The form of these Bianchi identitites in local coordinates is 
\begin{align}
&  R^d{}_{[abc]}=\nabla^{\E}_{[a}T_{bc]}{}^d+T_{e[a}{}^dT_{bc]}{}^e+\rho^{d\mu} S_{ abc\mu}\,,\\[4pt]
& \big(\nabla^{\E}_{[c|} R^e{}_{d|ab]}+R^e{}_{df[c}T_{ab]}{}^{f}\big)e^d=-\rho^{d\mu} S_{ abc\mu}\nabla^{\E}_{d}e^e\,,
\end{align}
where in both equations we have skew symmetrization in indices $a$, $b$ and $c$. We observe that, save the appearance of the basic curvature tensor, these Bianchi identities have the same form as in ordinary differential geometry. In the present case, the existence of a 3-bracket (which is absent for Lie algebroids) introduces the modification that contains the basic curvature tensor. 

To write the formulas for the components of the torsion and curvature tensors in local coordinates, first we note that the bracket $\ell_2$ is given as 
\begin{equation}
    \ell_{ab}{}^{c}=C_{ab}{}^{c}-\eta^{cd}\rho_{d}{}^{\mu}\Gamma_{\mu ab}\,.
\end{equation}
Then we obtain 
\begin{eqnarray}
    T_{ab}{}^{c}&=&\omega_{ab}{}^{c}-\omega_{ba}{}^{c}-C_{ab}{}^{c}+\eta^{cd}\rho_{d}{}^{\mu}\Gamma_{\mu ab}\,,\\[4pt]
    R^{a}{}_{bcd}&=&2\rho_{[c}{}^{\mu}\partial_{\mu}\omega_{d]b}{}^{a}+2\omega_{[c| f}{}^{a}\omega_{|d]b}{}^{f} \nonumber \\[4pt] && -\,(C_{cd}{}^{f}-\eta^{fg}\rho_{g}{}^{\mu}\Gamma_{\mu cd})\,\omega_{fb}{}^{a}\,, 
    \label{curvature local}
\end{eqnarray}
where skew symmetrization in the latter refers only to the indices $c$ and $d$. Thus, the curvature tensor is skew in the last two indices. It is also skew in the first two if the connection is $\eta$-metric. 
However, even if the connection is $\eta$-metric, the curvature tensor is not symmetric under the exchange of the pairs of indices. In particular, in that case we have
\begin{eqnarray}
&& \langle R_{\scriptscriptstyle\nabla}^{\scriptscriptstyle\nabla^{\E}}(e_1,e_2)e_3,e_4\rangle
    -\langle R_{\scriptscriptstyle\nabla}^{\scriptscriptstyle\nabla^{\E}}(e_3,e_4)e_1,e_2\rangle = \nonumber \\[4pt] &&\hspace{1cm}\big\langle\big(\dd^{\scriptscriptstyle\nabla^{\E}}T_{\scriptscriptstyle\nabla}^{\scriptscriptstyle\nabla^{\E}}+\ell_1\ell_3\big)(e_1,e_2,e_{[3}),e_{4]}\big\rangle\nonumber \\[4pt] &&\hspace{0.8cm}-\big\langle\big(\dd^{\scriptscriptstyle\nabla^{\E}}T_{\scriptscriptstyle\nabla}^{\scriptscriptstyle\nabla^{\E}}+\ell_1\ell_3\big)(e_3,e_4,e_{[1}),e_{2]}\big\rangle\,.
\end{eqnarray}
Even for the generalised connection with vanishing torsion we have a contribution due to homotopy Jacobi identity \eqref{Linf relations 2} for the binary bracket $\ell_2$.

\subsection{Generalised metric \& Koszul formula}

We would now like to move toward gravity. Up to this point, we have not introduced any metric that would serve as a dynamical object ($\eta$ is a background metric in this respect). Consider now a Riemannian metric $g=(g_{\mu\nu})$ on $M_0$, with Levi-Civita connection denoted as~$\mathring\nabla$, and a generalised metric $\mathcal H=(\mathcal H_{ab})$ on the Courant algebroid. The latter is defined as a maximal positive definite subbundle $V_+$ of $E$ with respect to $\eta$ or, equivalently, as an orthogonal endomorphism $\tau\in \text{End}(E)$, $\tau^2=\text{id}$, with $\mathcal H(e,e')=\eta(e,\tau(e'))$. There are now two additional compatibility conditions that one could possibly implement, namely $\mathcal H$-metricity conditions for ordinary and generalised connections: 
\begin{equation}
    \nabla\mathcal H=0 \quad \text{and} \quad \nabla^{\E}\,\mathcal H=0\,.
\end{equation}
One may now ask whether there exists a unique generalised connection $\nabla^{\E}$ which is generalised metric compatible and torsion free (with reference to the torsion defined in \eqref{torsion}). In the Lie 2-algebroid framework described in the previous section, the answer is affirmative for a \emph{fixed} $TM_0$-on-$E$ connection $\nabla$. There exists an analog of the Koszul identity of Riemannian geometry~\eqref{genkoszul}, which reads
\begin{align}  
&\mathcal H(e'',\nabla^{\E}_{e}e')= \nonumber \\[4pt]&  
\tfrac 12 \big(\rho(e')\mathcal H(e'',e) + \rho(e)\mathcal H(e',e'')-\rho(e'')\mathcal H(e',e) \label{Koszul} \\[4pt] 
& +\mathcal H\big(e'',\ell_2(e,e')\big) 
- \mathcal H\big(e',\ell_2(e,e'')\big)
-\mathcal H\big(\ell_2(e',e''),e\big)\big)\,.\nonumber
\end{align}  
This is the Koszul identity in generalised Riemannian geometry. The coefficients of the unique connection obtained that way are found to be 
\begin{eqnarray}
    \omega_{ab}{}^{d}&=&\tfrac 12 \mathcal H^{dc}\big(
    \rho_{b}{}^{\mu}\partial_{\mu}\mathcal H_{ac}
    +\rho_{a}{}^{\mu}\partial_{\mu}\mathcal H_{bc}
    -\rho_{c}{}^{\mu}\partial_{\mu}\mathcal H_{ab} \nonumber\\[4pt] && \hspace{35pt} +\mathcal H_{ce}C_{ab}{}^{e}-\mathcal H_{be}C_{ac}{}^{e}-\mathcal H_{ae}C_{bc}{}^{e} \nonumber\\[4pt] && +\eta^{ef}\rho_{f}{}^{\mu}(-\mathcal H_{ce}\Gamma_{\mu ab}+\mathcal H_{be}\Gamma_{\mu ac}+\mathcal H_{ae}\Gamma_{\mu bc})\big)\,.\qquad  \label{Koszul local}
\end{eqnarray}
We may call this the \emph{Levi-Civita generalised connection} for the Lie 2-algebroid associated to the split QP2 manifold via a given $\nabla$.

The following remarks are in order. As in ordinary Riemannian geometry, the Koszul identity admits a direct generalization for any fixed torsion by replacing $\ell_2\mapsto \ell_2+T^{\scriptscriptstyle\nabla^{\E}}$ in \eqref{Koszul}.  This means that we may also ask for a generalised metric compatible connection with a \emph{given} torsion, not necessarily vanishing, and the unique answer is obtained by the replacement $C_{ab}{}^{c}\mapsto C_{ab}{}^{c}+T_{ab}{}^{c}$ in Eq. \eqref{Koszul local}. Moreover, we note that the above identity reduces to the one for the Levi-Civita connection on a Lie algebroid \cite{BOUCETTA201157} when restricted to Dirac structures, i.e. maximal isotropic and involutive subbundles of $E$. Isotropy implies that the third line in \eqref{Koszul local} vanishes, and involutivity reduces it to Eq.\,(9) of \cite{BOUCETTA201157}. Finally, the generalised Koszul identity reduces to the usual one for Riemannian geometry when the Dirac structure is the canonical Lie algebroid on the tangent bundle $TM_0$, with anchor the identity map and bracket the commutator of vector fields. Then \eqref{Koszul local} becomes  
\begin{equation}
    \omega_{ab}{}^{c}|_{TM_0}=\omega_{\mu\nu}{}^{\rho}=\tfrac 12 g^{\rho\sigma}(\partial_{\mu}g_{\nu\sigma}+\partial_{\nu}g_{\mu\sigma}-\partial_{\sigma}g_{\mu\nu})\,,
\end{equation}
which is the Christoffel connection.  

\subsection{Comparison to other approaches}
\label{sec: comparison}

This approach to defining torsion and curvature for Courant algebroids and the Koszul identity for a generalised connection has its roots in \cite{Boffo:2019zus}. The new element here is that the $\ell_2$ bracket is not chosen arbitrarily, instead it follows from a systematic construction and it is part of a complete structure, the Lie 2-algebroid, which automatically yields Bianchi identities too. 

On the contrary, previous approaches either dispense with a well-defined, covariant Riemann tensor and use only an unambiguously defined Ricci tensor \cite{Coimbra:2011nw,Coimbra:2012yy,Severa:2018pag,Cavalcanti:2024uky,Garcia-Fernandez:2020ope} or define torsion and curvature differently \cite{10.1093/acprof:oso/9780199534920.003.0018,Jurco:2016emw,Aschieri:2019qku,Cueca:2019fej}. In~\cite{10.1093/acprof:oso/9780199534920.003.0018}, the definition of torsion of a generalised connection is~\footnote{The definition in \cite{10.1093/acprof:oso/9780199534920.003.0018} is written for the antisymmetrised Courant bracket. This is the version that corresponds to the Dorfman bracket.} 
\begin{equation}
T_{{\scriptscriptstyle{\text{G}}}}^{\scriptscriptstyle\nabla^{\E}}(e,e',e'')=\eta(\nabla^{\E}_ee'-\nabla^{\E}_{e'}e-[e, e'],e'')+\eta(\nabla_{e''}^{\E}e,e')\,.   \label{Gualtieri torsion} 
\end{equation}
We will call this the \emph{Gualtieri tensor} to avoid confusion with our definition.
Note that there exists a distinguished generalised connection, the $\dot\nabla^{\E}$ defined in \eqref{nabla dot}, for which~\eqref{torsion} and \eqref{Gualtieri torsion} are identical. For any other generalised connection, the two definitions differ by an endomorphism-valued $E$-section $s$, 
\begin{equation}
T_{\scriptscriptstyle{\text{G}}}^{\scriptscriptstyle\nabla^{\E}}-T^{\scriptscriptstyle\nabla^{\E}}_{\scriptscriptstyle{\nabla}}=s\,,
\end{equation}
with $s=\nabla^{\E}-\dot\nabla^{\E}$. Using \eqref{Gualtieri torsion}, a Levi-Civita generalised connection is not unique, due to undetermined parts of the connection \cite{Coimbra:2011nw}; rather, there is an infinite number of such connections in general \cite{Jurco:2016emw}.  Our construction ameliorates this unsatisfactory feature at the expense of a choice of ordinary connection $\nabla$. We note that there have been other instances of Levi-Civita type connections defined with the help of the Gualtieri tensor~\cite{Tipler2016mat, Hu2024}.

As for the Riemann curvature tensor, a proposal to improve the other unsatisfactory feature of its lack of covariance appeared in the context of doubled geometry in \cite{Hohm:2011si,Hohm:2012mf}, whose incarnation in generalised geometry is given in \cite{Jurco:2016emw} as 
\begin{eqnarray}
    {\mathcal R}^{\scriptscriptstyle\nabla^{\E}}(\hat{e},\hat{e}',e,e')&=& \tfrac 12 \big(R^{\scriptscriptstyle\nabla^{\E}}_0(\hat{e},\hat{e}',e,e')+R^{\scriptscriptstyle\nabla^{\E}}_0(e',e,\hat{e}',\hat{e})\big) \nonumber\\[4pt] &&+\,\eta\big(K(e,e'),K(\hat{e},\hat{e}')\big)\,,\label{curvature other}
\end{eqnarray}
where $R_0^{\scriptscriptstyle\nabla^{\E}}(e,e')=[\nabla^{\E}_{e},\nabla^{\E}_{e'}]-\nabla^{\E}_{[e,e']}$ and  $\eta(K(e,e'),\hat{e})=\eta(\nabla^{\E}_{\hat{e}}e,e')$. For the induced generalised connection $\dot\nabla^{\E}$, all definitions agree, $R_{\scriptscriptstyle{\nabla}}^{\scriptscriptstyle{\dot\nabla}^{\E}}=\mathcal{R}^{\scriptscriptstyle{\dot\nabla}^{\E}}=R_0^{\scriptscriptstyle{\dot\nabla}^{\E}}$ (note that the naive curvature $R_0$ \emph{is} covariant for this and only this generalised connection. This is no longer true for other generalised connections). 

\section{Gravity from the Lie 2-algebroid}

\subsection{Generalised gravity action} 

We would now like to use the construction outlined in the previous sections to write down models for gravity from generalised geometry in the (generalised) metric formalism. 
To recapitulate the logic, we start with a QP2 manifold $M_2$ and split it to a graded vector bundle $T^{\ast}[2]M_0\oplus E[1]$ with a connection $\nabla$ which is compatible with the inner metric. This results in a bracket $\ell_2$ with the right properties to unambiguously define generalised torsion and curvature  for generalised connections. Introducing a generalised metric and requiring compatibility with a generalised connection of given torsion, fixes the latter through the Koszul formula. One may then compute the Riemann curvature tensor for the obtained generalised connection. What remains is to define the generalised Ricci tensor and Ricci scalar. 

The minimal way to define a generalised Ricci tensor is to contract the first and third entries of the Riemann tensor, as in ordinary Riemannian geometry: 
\begin{equation}
    R_{ab}:=R^{c}{}_{acb}\,,
\end{equation}
with reference to Eq.~\eqref{curvature local}. We note that $R_{ab}$ is not necessarily symmetric. In particular, we have
\begin{equation}
    2R_{[ab]}=\nabla^{\E}_{[a}T_{cb]}{}^c+T_{e[a}{}^cT_{cb]}{}^e+\eta^{ce}\rho_{e}{}^\mu S_{ acb\mu}\,,
\end{equation}
where the last term comes from the homotopy Jacobi identity~\eqref{Linf relations 2} for the bracket. The appearance of torsion in the antisymmetric part of the Ricci tensor is a feature of ordinary differential geometry too.
One may define a symmetrised version, as in \eqref{curvature other} for the Riemann tensor of~\cite{Jurco:2016emw}, however this will not be necessary for our purposes. 

There are in principle two ways to form a curvature scalar, by taking a trace with respect to (the inverse of) each of the two (nondegenerate) inner products $\eta$ and~$\mathcal H$~\cite{Jurco:2016emw}. We denote these curvature scalars as $R_{\eta}$ and $R_{\mathcal H}$: 
\begin{equation}
    R_{\eta}:=\eta^{ab}R_{ab} \quad \text{and} \quad R_{\mathcal{H}}:=\mathcal{H}^{ab}R_{ab}\,.
\end{equation}
Mathematically, both are admissible to construct an action functional, however $R_{\mathcal{H}}$ is more natural from the point of view of general relativity, since it is formed with the dynamical metric $\mathcal{H}$ instead of the nondynamical background metric $\eta$. In fact, it will turn out that $R_{\eta}$ vanishes in the case of the closed string effective action that we will consider. Therefore, as a minimal model we 
write down a generalised analogue of the Einstein-Hilbert action for a given Lie 2-algebroid, subject to a choice of volume form $\omega$: 
\begin{equation}
    S_{\text{Lie2oid}}=\int \omega R_{\mathcal H}\,.\label{action general}
\end{equation}
With some physical input, we select the following volume form on the manifold: 
\begin{equation}\label{eq: volume form}
    \omega=e^{-2\phi}\sqrt{-g}\,\dd^{D}x\,,
\end{equation}
where $D$ is the number of spacetime dimensions, differing from the usual one in Riemannian geometry by a conformal factor. Note that $e^{-2 d}:=e^{-2\phi}\sqrt{-g}$ is often called the T-duality invariant dilaton due to its good transformation property under $O(D,D)$ transformations \cite{Hull:2006va}.
 A~similar action in Palatini form, but for the different definition of Riemann and Ricci curvatures \eqref{curvature other}, was written in \cite{Jurco:2022nlr}. 
 As it stands, the action \eqref{action general} is generic and we will not explore it in full generality here. Instead our goal is to investigate under which geometric conditions it results in the low energy effective action for the massless common sector of closed string theory.

Before we proceed, some remarks about the volume form are due. Choosing a volume form amounts to a trivialization of the determinant line bundle $\mathrm{det}(T^{\ast}M_0)$ through a section $\omega: M_0\to \mathrm{det}(T^{\ast}M_0)$. This induces an isomorphism $E\cong E\otimes \mathrm{det}(T^{\ast}M_0)$, which is precisely the approach established in~\cite{Coimbra:2011nw}. In ordinary differential geometry, the volume form allows to define the divergence of a vector field as $\mathrm{div}_{\omega}(X) \, \omega = L_{X}\omega = \dd\iota_{X}\omega$. Similarly, in the present context the divergence induced by the volume form $\omega$ is given through the anchor map as 
\begin{equation}
\mathrm{div}_{\omega}(e)\,\omega=L_{\rho(e)}\omega\,,
\end{equation}
where $e\in\Gamma(E)$. Furthermore, any generalised connection $\nabla^{\E}$ can be used to define a divergence operator as follows \cite{Garcia-Fernandez:2020ope, Jurco:2016emw}: 
 \begin{equation}
     \mathrm{div}_{\nabla^{\E}}(e) = \mathrm{Tr}(\nabla^{\E}e)\,. 
 \end{equation}
 For ordinary connections on Courant algebroids, we may define the divergence to be the one corresponding to the canonically induced generalised connection through the anchor, namely $\mathrm{div}_{\dot{\nabla}^{\E}}(e)$. 
 
 When we consider both a generalised connection and a volume form, it is meaningful to ask whether the two divergence operators induced by them are equal. Hence we will call a generalised connection \emph{$\omega$-unimodular} when its divergence is equal to the one defined by the volume form $\omega$, i.e. when
 \begin{equation}
 \mathrm{div}_{\nabla^{\E}}(e)=\mathrm{div}_{\omega}(e)\,.
 \end{equation}
 Such unimodular connections are also important in the approach of Refs.\,\cite{Jurco:2016emw,Jurco:2022nlr}. We will call an ordinary $TM$-on-$E$ connection -- say $\nabla$ -- \emph{$\omega$-unimodular} when the induced connection $\dot{\nabla}^{\E}$ is $\omega$-unimodular. Unimodularity implies flatness of the induced connection on the determinant bundle $\mathrm{det}(T^{\ast}M_0)$, however, it is not implied by and does not imply absence of torsion. Nevertheless, it is important to note that for the choice of volume form~\eqref{eq: volume form} unimodular generalised connections are never symmetric; there always exists a $\phi$-dependent contribution to the trace of their torsion tensor. 
 

\subsection{Properties of the standard Courant algebroid}

We now focus on the case that corresponds to the $H$-twisted standard -- or exact -- Courant algebroids, classified by the \v{S}evera class (see \cite{Garcia-Fernandez:2020ope} for a self-contained presentation), which is the geometric structure encountered in generalised geometry. This uses the generalised tangent bundle $E=TM_0\oplus T^{\ast}M_0:=\mathbb{T}M_0$, with sections comprising pairs of vector fields and 1-forms $e=(X,\lambda)\in\Gamma(E)$ and anchor the projection to $TM_0$, $\rho(e)=X$. The inner metric is the $O(D,D)$ invariant one: 
\begin{equation}
    \eta(e,e')=\lambda(X')+\lambda'(X)\,, 
    \quad \eta=\begin{pmatrix}
        0 & 1 \\ 1 & 0
    \end{pmatrix}\,,
\end{equation}
so that the fiber isomorphism $\eta^\sharp:\mathbb{T}M_0\to \mathbb{T}M_0$ is the identity map.
The $H$-twisted Dorfman bracket is given as 
\begin{equation}
    [e,e']_H= \big([X,X']_{\text{Lie}},\mathcal L_{X}\lambda'-\iota_{X'}\dd\lambda-H(X,X',-)\big)\,,
\end{equation}
where $H$ is a closed 3-form, which locally is obtained as $H=\dd B$ for the Kalb-Ramond 2-form field $B$. 

In graded-geometric terms, the standard Courant algebroid corresponds to the QP2 manifold $M_2=T^{\ast}[2]T[1]M$ with Hamiltonian as in Eq.~\eqref{Hamiltonian QP2}, in which $\rho_{a}{}^{\mu}=(\rho_{\nu}{}^{\mu},\rho^{\nu\mu})=(\delta_{\nu}{}^{\mu},0)$ and $C_{abc}$ having $C_{\mu\nu\rho}=-H_{\mu\nu\rho}$ as its only nonvanishing component. The Maurer-Cartan equation is identically satisfied given these data, due to the closedness of the 3-form $H$. In the spirit of our earlier discussion, we consider a connection $\nabla$ that splits $M_2$ and yields a Lie 2-algebroid with a skew symmetric bracket $\ell_2$ as in Eq.~\eqref{ell 2}. 
Recall that for this to work, we must assume that $\nabla\eta=0$, which translates into $\Gamma_{\mu ab}=-\Gamma_{\mu ba}$ for the connection coefficients of $\nabla$. In the present case we can write them as 
\begin{equation}
    \Gamma_{\mu a}{}^{b}=\begin{pmatrix}
        \Gamma_{\mu\nu}{}^{\rho} & \Gamma_{\mu\nu\rho} \\ \Gamma_{\mu}{}^{\nu\rho} & \Gamma_{\mu}{}^{\nu}{}_{\rho}
    \end{pmatrix}\,; \label{gamma matrix}
\end{equation}
$\eta$-metricity then tells us that $\Gamma_{\mu\nu}{}^{\rho}=-\Gamma_{\mu}{}^{\rho}{}_{\nu}$ and $\Gamma_{\mu(\nu\rho)}=0=\Gamma_{\mu}{}^{(\nu\rho)}$. 
Note that the vector part of $\ell_2$ vanishes identically, 
\begin{equation}
    \ell_{ab}{}^{\mu}=0\,.
\end{equation}
On the other hand, its 1-form part depends on the connection coefficients and $H$, with nonvanishing components 
\begin{subequations}
    \begin{align}
    \ell_{\mu\nu\rho}&=-(H_{\mu\nu\rho}+\Gamma_{\mu\nu\rho})\,, \\[4pt] 
    \ell^{\mu}{}_{\nu\rho}&=-\ell_{\nu}{}^{\mu}{}_{\rho}=\Gamma_{\rho\nu}{}^{\mu}\,,\\[4pt] 
    \ell^{\mu\nu}{}_{\rho}&=-\Gamma_{\rho}{}^{\mu\nu}\,.
\end{align}
 \end{subequations}
 One can fully determine the bracket, once a connection~$\nabla$ is fixed, and we will do so from meaningful geometric constraints in the next section.

There is a distinguished class of automorphisms of the vector bundle $\mathbb{T}M_0$ under which the Dorfman bracket transforms covariantly, namely: the $B$-transforms \cite{gualtieri2011generalized}. For any differential 2-form $B$ on $M_0$, we define the $B$-transform of an element $(X,\lambda)$ as $e^B(X,\lambda)=(X,\lambda+\iota_XB)$. The image of the transformed element through the anchor map is the same as the original one. The Dorfman bracket then transforms covariantly with respect to the $B$-transform, that is to say:
\begin{equation}
    e^B\left([e^{-B}e,e^{-B}e']_H\right)=[e,e']_{H+dB}~.
\end{equation}
The connection $\nabla$ transforms as well; we define the connection $\nabla^{\B}:\Gamma(TM)\times\Gamma(E)\to \Gamma(E)$ as its $B$-transform 
\begin{equation}
\nabla^{\B}_X(e)=e^B\left(\nabla_X(e^{-B}e)\right)\,,
\end{equation}
i.e. $\nabla^{\B}=e^{B}\circ\nabla\circ e^{-B}$.
Although the action of the $B$-transform on the vector field $X$ is not zero, the resulting form component $\iota_XB$ does not entail any action of the connection.
From these definitions, we observe that the 2-bracket and 3-bracket of the Lie 2-algebroid transform covariantly under B-transforms (for the sake of clarity, we label the brackets by their constitutive elements):
\begin{align}
    e^{B}\left(\ell_2^{\H,\nab}(e^{-B}e,e^{-B}e')\right)&=\ell_2^{\HB,\nab^{\B}}(e,e')\,,\nonumber\\[4pt]
    e^{B}\left(\ell_2^{\H,\nab}(e^{-B}e,e^{-B}\lambda)\right)&=\ell_2^{\HB,\nab^{\B}}(e,\lambda)\,,\nonumber\\[4pt]
    e^{B}\left(\ell_3^{\H,\nab}(e^{-B}e,e^{-B}e',e^{-B}e'')\right)&=\ell_3^{\HB,\nab^{\B}}(e,e',e'')\,.
\end{align}
Notice that the $B$-transform acts as the identity on the cotangent bundle, giving rise to the last two formulas. From this we deduce that a $B$-transform is a \emph{strict} Lie 2-algebroid automorphism.

Additionally, the $B$-transform, by acting on a generalised connection $\nabla^{\E}:\Gamma(E)\times\Gamma(E)\to\Gamma(E)$, defines a new generalised connection as follows:
\begin{equation}    \nabla^{\E,\B}_e(e')=e^B\left(\nabla^{\E}_{e^{-B}e}(e^{-B}e')\right)\,.
\end{equation}
From this we observe that the torsion and the curvature transform covariantly:
\begin{align}
e^B\left(T_{\H,\scriptscriptstyle\nabla}^{\scriptscriptstyle\nabla^{\E}}(e^{-B}e,e^{-B}e')\right)&=T_{\HB,\scriptscriptstyle\nabla^{\B}}^{\scriptscriptstyle\nabla^{\E,\B}}\,,\nonumber \\[4pt]
e^B\left(R_{\H,\scriptscriptstyle\nabla}^{\scriptscriptstyle\nabla^{\E}}(e^{-B}e,e^{-B}e')\right)&= R_{\HB,\scriptscriptstyle\nabla^{\B}}^{\scriptscriptstyle\nabla^{\E,\B}}\,.
\end{align}
In particular, we observe that if a generalised connection~$\nabla^{\E}$ is torsion (resp. curvature) free, then is $B$-transform is also torsion (resp. curvature) free in the transformed frame. From this we deduce that the $B$-transform of the generalised Levi-Civita connection with respect to a given torsion tensor  and a given generalised metric is the generalised Levi-Civita connection with respect to the $B$-transformed torsion tensor and generalised metric (described below).

\subsection{Geometric conditions for a choice of $\nabla$}

Next, we introduce a generalised metric and we consider a frame where it takes the simple form 
\begin{equation}
    \mathcal H=\begin{pmatrix}
        g & 0 \\ 0 & g^{-1}
    \end{pmatrix}\,;
\end{equation}
the general form is obtained by $B$-transforms as 
\begin{equation}
    \mathcal H^{\B}= \begin{pmatrix}
        1 & B \\ 0 & 1
    \end{pmatrix}\mathcal H \begin{pmatrix}
        1 & 0 \\ -B & 1
    \end{pmatrix}= \begin{pmatrix}
        g-Bg^{-1}B & Bg^{-1} \\ -g^{-1}B & g^{-1}
    \end{pmatrix}\,. 
\end{equation}
A choice of generalised metric results in a splitting of $\mathbb{T}M_0$ in eigenbundles. This is so because the operator $\eta^{-1}\mathcal{H}^{\B}$ satisfies $(\eta^{-1}\mathcal H^{\B})^2=1$ and therefore there are two projectors 
\begin{equation}
    P_{\pm}:=\tfrac 12(1\pm\eta^{-1}\mathcal{H}^{\B})\,, 
\end{equation}
which project on the corresponding eigenbundles. They satisfy $P_{\pm}^{2}=P_{\pm}$ and $P_+P_-=0=P_-P_+$. The splitting of $\mathbb{T}M_0$ due to the generalised metric is independent from the splitting of $M_2$ due to the ordinary connection $\nabla$ introduced earlier. Since both structures are important in our construction, it makes sense to require compatibility between them, in the sense that the connection $\nabla$ preserves the eigenbundle decomposition due to the generalised metric $\mathcal H$. This translates to the condition
\begin{equation}
    P_{\mp}\left(\nabla_{X}\left(P_{\pm}(e)\right)\right)=0\,,
\end{equation}
or, in operator language, $P_{\mp}\circ\nabla\circ P_{\pm}=0$. This requirement is equivalent to the $\mathcal H$-metricity of $\nabla$, $\nabla\mathcal{H}=0$, which we impose from now on. Once this is true in the diagonal frame, it remains true in any frame due to the behavior under $B$-transforms. At the level of connection components, $\mathcal{H}$-metricity imposes in addition that 
\begin{equation}\label{eq:Hmetricity}
    (\Gamma-\mathring\Gamma)_{\mu(\nu}{}^{\rho}g_{\sigma)\rho}=0 \quad \text{and} \quad \Gamma_{\mu\nu\rho}=\Gamma_{\mu}{}^{\kappa\lambda}g_{\kappa\nu}g_{\lambda\rho}\,,
\end{equation}
where we recall that $\mathring{\Gamma}_{\mu\nu}{}^{\rho}$ are the coefficients of the Levi-Civita connection for the Riemannian metric $g$. This still does not completely fix the connection coefficients of $\nabla$. 

As a final step, we would like to impose a torsion condition on the canonical induced connection $\dot\nabla^{\E}$ for $\nabla$ given by (\ref{nabla dot}). Since this is a generalised connection, its torsion tensor is given by Eq.~\eqref{torsion}.
We denote the contracted tensor as $T^{\scriptscriptstyle{\dot\nabla^{\E}}}_{\scriptscriptstyle{\nabla}}(e,e',e''):=\eta(T^{\scriptscriptstyle{\dot\nabla^{\E}}}_{\scriptscriptstyle{\nabla}}(e,e'),e'')$.  Note that this contracted tensor is completely skew symmetric~{\footnote{It is worth mentioning that once $\eta$-metricity of a generalised connection is imposed, the Gualtieri tensor is always completely skew symmetric. This is not always the case for the torsion tensor we have defined in this paper. However, it is true for the induced connection, since in that case the two definitions agree, as already mentioned in Section \ref{sec: comparison}.}}. 

The vector-vector sector of the torsion tensor is
\begin{align}
     T^{\scriptscriptstyle{\dot\nabla}^E}_{\scriptscriptstyle{\nabla}}(\partial_\mu&,\partial_\nu)
     =\Big(\Gamma_{\mu\nu}{}^{\rho}-\Gamma_{\nu\mu}{}^{\rho}\Big)\partial_\rho\label{eq:vecvecsector}\\
     &+\Big(\Gamma_{\mu\nu\rho}-\Gamma_{\nu\mu\rho}+\left\langle \nabla_{\partial_\rho}(\partial_\mu),\partial_\nu\right\rangle+H_{\mu\nu\rho}\Big)dx^\rho\nonumber\,.
\end{align}
The first component is the torsion of the $\Gamma_{\mu\nu}{}^{\rho}$ sector of the connection $\nabla$, while the second component reads:
\begin{equation}
    T^{\scriptscriptstyle{\dot\nabla}^E}_{\scriptscriptstyle{\nabla}}(\partial_\mu,\partial_\nu,\partial_\rho)=\Gamma_{\mu\nu\rho}-\Gamma_{\nu\mu\rho}+\Gamma_{\rho\mu\nu}+H_{\mu\nu\rho}\,.
\end{equation}
Therefore, fixing a value for the vector-vector sector \eqref{eq:vecvecsector} of the torsion will fix the components $\Gamma_{\mu\nu}{}^\rho$ and $\Gamma_{\mu\nu\rho}$.
The remaining sectors of the torsion are given by:
\begin{subequations}
    \begin{eqnarray}\label{torsion dotfixed}
&&T^{\scriptscriptstyle{\dot\nabla}^E}_{\scriptscriptstyle\nabla}(\partial_\mu,dx^\kappa,dx^\lambda)=\left\langle\nabla_{\partial_\mu}(dx^\kappa),dx^\lambda\right\rangle={\Gamma_{\mu}{}^{\kappa\lambda}}
    \,,\\[4pt]
&&T^{\scriptscriptstyle{\dot\nabla}^E}_{\scriptscriptstyle\nabla}(dx^\kappa,dx^\lambda,dx^\mu)=0\,,
\end{eqnarray}
\end{subequations}
and permutations thereof due to skew symmetry.
The first 
is fixed from $\Gamma_{\mu\nu\rho}$
by $\mathcal{H}$-metricity, Eq. \eqref{eq:Hmetricity}, while the 
second one vanishes by construction.

Furthermore, since we have already trivialised the determinant bundle by choosing the volume form $\omega$ in Eq.\,\eqref{eq: volume form}, it is natural to assume that $\nabla$ is compatible with this trivialization. In other words, we can choose it to be an $\omega$-unimodular connection. As already mentioned, this implies that the torsion of the induced generalised connection cannot vanish. Instead we have the following torsion condition :
\begin{equation}
    T^{\scriptscriptstyle{\dot\nabla}^E}_{\scriptscriptstyle{\nabla}}(\partial_\mu,\partial_\nu)=\tfrac{2}{D-1}
\,d\phi\wedge dx^\rho(\partial_\mu,\partial_\nu) \,\partial_\rho\,,
\end{equation}
which decomposes into the following two conditions: 
\begin{subequations}
    \begin{eqnarray}\label{torsion dot}
 &&T^{\scriptscriptstyle{\dot\nabla^{\E}}}_{\scriptscriptstyle{\nabla}}(\partial_\mu,\partial_\nu,dx^\rho)=\tfrac{2}{D-1}
\,\dd\phi\wedge dx^\rho(\partial_\mu,\partial_\nu) \,,  
\label{bullet torsion cndition 1}\\[4pt]
&& T^{\scriptscriptstyle{\dot\nabla^{\E}}}_{\scriptscriptstyle{\nabla}}(\partial_\mu,\partial_\nu,\partial_\rho)= 0\,,
\label{bullet torsion condition 2}
\end{eqnarray}
\end{subequations}
where $\phi$ is going to be identified with the dilaton. 

All together, the above requirements completely fix the connection $\nabla$. Equivalently, there exists a unique  Lie 2-algebroid corresponding to a split QP2 manifold $M_2$ which is obtained by means of a connection $\nabla$ which is $\eta$-compatible, $\mathcal H$-compatible, $\omega$-unimodular and its induced generalised connection satisfies $T_{\scriptscriptstyle{\nabla}}^{\scriptscriptstyle{\dot\nabla^{\E}}}(X,Y,Z)=0$. Its diagonal coefficients in the matrix form \eqref{gamma matrix} are 
fixed by~\eqref{bullet torsion cndition 1} ($\omega$-unimodularity) and $\eta$-metricity, and they are 
found to be
\begin{equation}
\Gamma_{\mu\nu}{}^{\rho}=-\Gamma_{\mu}{}^{\rho}{}_{\nu}=\mathring{\Gamma}_{\mu\nu}{}^{\rho}-\tfrac{2}{D-1}(\delta_{\mu}^{\rho}\partial_{\nu}\phi-g_{\mu\nu}\partial^{\rho}\phi)\,.    
\end{equation}
The off-diagonal coefficients are fixed by the second condition \eqref{bullet torsion condition 2} and they are completely skew symmetric:  
\begin{equation}
    \Gamma_{[\mu\nu\rho]}=-\tfrac 13 H_{\mu\nu\rho}\,. 
\end{equation}
The piece of the off-diagonal coefficients with $\Gamma_{[\mu\nu]\rho}=0$, namely the ``hook'' part, vanishes.
Note that in absence of the dilaton field, the diagonal components are identical to the Levi-Civita connection. 

For the sake of completeness, we calculate the components of the basic curvature tensor for this connection on the Courant algebroid. The independent components are 
\begin{eqnarray}
    &&S_{\alpha\beta\gamma\mu}=-\tfrac 23 \mathring\nabla_{\mu}H_{\alpha\beta\gamma}-\tfrac 4{D-1}(\delta_{\mu}^{\rho}\partial_{[\alpha}\phi-g_{\mu[\alpha}\partial^{\rho}\phi)H_{\beta\gamma]\rho}\,,\nonumber\\[4pt] 
    &&S^{\alpha\beta\gamma}{}_{\mu}=\tfrac 13 H_{\rho}{}^{[\beta\gamma}H_{\mu}{}^{\alpha]\rho}\,,\nonumber\\[4pt]
    &&S_{\alpha\beta}{}^{\gamma}{}_{\mu}=-\mathring{R}^{\gamma}{}_{\mu\alpha\beta}-\tfrac 29 H_{\rho\alpha\beta}H_{\mu}{}^{\gamma\rho} \nonumber\\[4pt] &&\quad\quad\quad +\tfrac 4{D-1}g_{\mu[\alpha}(\mathring\nabla_{\beta]}\partial^{\gamma}\phi+\tfrac 2{D-1}\partial_{\beta]}\phi\partial^{\gamma}\phi)\,,\nonumber\\[4pt] 
    &&S_{\alpha}{}^{\beta\gamma}{}_{\mu}=\tfrac 13 \mathring\nabla_{\alpha}H_{\mu}{}^{\beta\gamma} \nonumber \\[4pt] 
    &&\quad\quad\quad -\tfrac 2{3(D-1)}(H_{\mu}{}^{\beta\gamma}\partial_{\alpha}\phi+2H_{\mu\alpha}{}^{[\beta}\partial^{\gamma]}\phi+g_{\mu\alpha}H^{\nu\beta\gamma}\partial_{\nu}\phi)\,.\nonumber \\
\end{eqnarray}
All other components are obtained using the fact that $S_{abc\mu}$ is fully skew symmetric in its first three indices. We observe that the components of the basic curvature for this connection do not vanish. 

\subsection{Geometric conditions for closed string action}

Having found meaningful geometric conditions that fix the ordinary connection $\nabla$, we have a distinguished split QP2 manifold and we can now use the Koszul identity to compute the unique $\mathcal H$-metric generalised connection~$\nabla^{\E}$ for any fixed torsion. 
We can then compute its Riemann tensor, its Ricci tensor and its Ricci scalar $R_{\mathcal{H}}$ (and also~$R_{\eta}$). There are two ways to proceed, recalling that there is still one more condition due, the $\eta$-metricity of the generalised connection~$\nabla^{\E}$. To obtain the closed string effective action we can impose $\eta$-metricity of $\nabla^{\E}$, in which case we must fix a nonzero torsion $T^{\scriptscriptstyle{\nabla^{\E}}}_{\scriptscriptstyle{\nabla}}$. Alternatively, we can impose that $T^{\scriptscriptstyle{\nabla^{\E}}}_{\scriptscriptstyle{\nabla}}=0$, in which case~$\nabla^{\E}$ will not satisfy $\eta$-metricity \footnote{This is reminiscent of the different formulations of general relativity in terms of torsion or distorsion tensors, called teleparallel and symmetric teleparallel formulations \cite{BeltranJimenez:2019esp}.}. We proceed with the first option.

The requirement of $\eta$-metricity for $\nabla^{\E}$ imposes that its last two indices are skew symmetric as in Eq.~\eqref{eta metricity omega}. On the other hand, $\mathcal{H}$-metricity imposes the following condition:
\begin{equation}
     \omega_{ab}{}^c=\mathring\omega_{ab}{}^c+\mathcal{H}^{cd}\left(K_{dab}^{\E}+L_{dab}\right)\,,
\end{equation}
where  on the right-hand side we defined:
\begin{align}
    \mathring\omega_{ab}{}^c
    &=\tfrac 12 \mathcal H^{cd}\big(\rho_{a}{}^{\mu}\partial_{\mu}\mathcal H_{bd}+\rho_{b}{}^{\mu}\partial_{\mu}\mathcal H_{ad}-\rho_{d}{}^{\mu}\partial_{\mu}\mathcal H_{ab}\big)\,, \nonumber\\[4pt]
    K_{dab}^{\E}
    &=\tfrac 12 \big(T_{ab}{}^e\mathcal{H}_{de}-T_{ad}{}^e\mathcal{H}_{be}-T_{bd}{}^e\mathcal{H}_{ae}\big)\,,\nonumber\\[4pt]
    L_{dab}
    &=\tfrac 12 \big(\ell_{ab}{}^e\mathcal{H}_{de}-\ell_{ad}{}^e\mathcal{H}_{be}-\ell_{bd}{}^e\mathcal{H}_{ae}\big)\,.
\end{align}
In particular $K^{\E}:=K_{{\scriptscriptstyle\nabla}}^{\scriptscriptstyle\nabla^E}$ is the generalised contorsion tensor for $\nabla^{\E}$.
Furthermore, we require that the generalised connection is $\omega$-unimodular. This guarantees that we can perform integration by parts as
\begin{equation}
\int \omega\, f\,\text{div}_{\nabla^{\E}}(X+\lambda) = -\int\omega\, X(f)\,,  
\end{equation}
for any function $f(x)\in\mathcal{C}^\infty(M_0)$. As was the case for $\nabla$, unimodularity of an $\eta$-compatible generalised connection requires a nonvanishing torsion tensor.

To apply the Koszul formula and guided by unimodularity, we ask for a generalised connection whose torsion tensor is fixed to have the following components:
\begin{subequations}\label{omega torsion conditions}
\begin{align}
T^{\scriptscriptstyle{\nabla^{\E}}}_{\scriptscriptstyle{\nabla}}(X,Y,Z)&=0\,, \\[4pt] 
T^{\scriptscriptstyle{\nabla^{\E}}}_{\scriptscriptstyle{\nabla}}(X,Y,\lambda)&=\tfrac{3}{2(D-1)}
\, \dd\phi\wedge \lambda(X,Y)\,.
\end{align}
\end{subequations}
The other components of the torsion tensor are fixed from these two conditions, unimodularity and metric compatibility. Observe that the torsion in this case is not completely skew symmetric. 
We can write the two torsion conditions \eqref{omega torsion conditions} together in the following compact form:
\begin{equation}
T^{\scriptscriptstyle{\nabla^{\E}}}_{\scriptscriptstyle{\nabla}}(\partial_\mu,\partial_\nu)=\tfrac{3}{2(D-1)}
\,d\phi\wedge dx^\rho(\partial_\mu,\partial_\nu) \,\partial_\rho\,.
\end{equation}

The unique generalised connection with these properties has components: 
\begin{align}
    \omega_{\mu b}{}^{c}&=\begin{pmatrix}
        \omega_{\mu\nu}{}^{\rho} & \omega_{\mu\nu\rho} \\ \omega_{\mu}{}^{\nu\rho} & \omega_{\mu}{}^{\nu}{}_{\rho}
    \end{pmatrix}\,, \label{omega matrix 1} \\[4pt] 
    \omega^{\mu}{}_{b}{}^{c}&=\begin{pmatrix}
        \omega^{\mu}{}_{\nu}{}^{\rho} & \omega^{\mu}{}_{\nu\rho} \\[4pt] \omega^{\mu\nu\rho} & \omega^{\mu\nu}{}_{\rho}
    \end{pmatrix}\,, \label{omega matrix 2}
\end{align}
with the four independent ones being
\begin{subequations}\label{omega final}
    \begin{align}
    \omega_{\mu\nu}{}^{\rho}&=\mathring\Gamma_{\mu\nu}{}^{\rho}+\tfrac{3}{2(D-1)}(g_{\mu\nu}\partial^{\rho}\phi-\delta_{\mu}^{\rho}\partial_{\nu}\phi)\,, \\[4pt] 
    \omega_{\mu\nu\rho}&=-\tfrac 13 H_{\mu\nu\rho}\,, \\[4pt] 
    \omega^{\mu\nu}{}_{\rho}&=\tfrac 16 H^{\mu\nu}{}_{\rho}\,, \\[4pt] 
    \omega^{\mu\nu\rho}&=\tfrac 1{2(D-1)}(g^{\mu\nu}\partial^{\rho}\phi-g^{\mu\rho}\partial^{\nu}\phi)\,,
\end{align}
\end{subequations}
and the rest following from $\eta$- and $\mathcal{H}$-metricities. We may now compute the Ricci tensor, whose components are
\begin{subequations}\label{Ricci}
\begin{align}
    R_{\mu\nu}&=\mathring{R}_{\mu\nu}-\tfrac 16 H_{\mu\kappa\lambda}H_{\nu}{}^{\kappa\lambda}\nonumber\\[4pt]
    &+\tfrac{1}{2(D-1)}( 3g_{\mu\nu}\mathring\nabla_\rho\partial^\rho\phi-(7-4D)\mathring\nabla_{\mu}\partial_{\nu}\phi)\nonumber\\[4pt]
    &+\tfrac{1}{(D-1)^2}((5-3D)g_{\mu\nu}\partial_\rho\phi\partial^\rho\phi+2(D-2)\partial_\mu\phi\partial_\nu\phi)\,,\\[4pt] 
    R^{\mu\nu}&=\tfrac 1{12} H^{\mu\kappa\lambda}H^{\nu}{}_{\kappa\lambda}-\tfrac{1}{D-1}g^{\mu\nu}\partial_\rho\phi\partial^\rho\phi\nonumber\\[4pt]
    &\qquad +\tfrac{1}{2(D-1)}(g^{\mu\nu}\mathring\nabla_\rho\partial^\rho\phi-\mathring\nabla^\mu\partial^\nu\phi)\,,
\end{align}
\end{subequations}
where $\mathring{R}_{\mu\nu}$ is the Ricci tensor for the Levi-Civita connection on $M_0$. The computation of the Ricci scalar $R_{\mathcal{H}}=g^{\mu\nu}R_{\mu\nu}+g_{\mu\nu}R^{\mu\nu}$ is straightforward and when it is inserted into the action \eqref{action general} yields,
\begin{eqnarray}
 S_{\text{Lie2oid}}&=&\int \dd^Dx \sqrt{-g} \,e^{-2\phi}\times 
 \nonumber\\[4pt] 
 &&\times\left(\mathring{R}-\tfrac 1{12}H^2-4\mathring\nabla_{\mu}\phi\mathring\nabla^{\mu}\phi+4\,\Box_{g}\phi\right) \nonumber \\[4pt]
 &=& S_{\text{closed string}}\,,
\end{eqnarray}
where $H^2=H_{\mu\nu\rho}H^{\mu\nu\rho}$ and $\Box_{g}$ is the Laplace-Beltrami operator. 
This is precisely the low energy effective action of the common sector in closed string theory, up to a total derivative, with $D=26$ for bosonic strings and $D=10$ for type II strings. Evidently, criticality cannot be accounted for in this purely geometric manner. We note that for the given generalised connection \eqref{omega final}, the components of the Ricci tensor in Eq.\,\eqref{Ricci} are symmetric. The other two components are not, $R_\mu{}^\nu\neq R^\nu{}_\mu$, but  \begin{equation}
    R_{\eta}=0
\end{equation} 
holds.
The same happens in Ref.\cite{Jurco:2022nlr}, where the generalised connection is different. Furthermore, for the given generalised connection \eqref{omega final},
the Gualtieri tensor \eqref{Gualtieri torsion} is not zero, with the only non-vanishing component being
\begin{equation}
T^{\scriptscriptstyle{\nabla^{\E}}}_{\scriptscriptstyle{\text{G}}}(\partial_\mu,\partial_\nu,dx^\rho)=\tfrac{1}{2(D-1)}
\,\dd\phi\wedge dx^\rho(\partial_\mu,\partial_\nu)~.
\end{equation}

We close this section with an important remark. Recall that we argued that the choice of ordinary connection is essentially immaterial, in the sense that any choice leads to the same Lie 2-algebroid up to quasi-isomorphism. However, when we constructed the gravity action in this section we completely fixed this connection, effectively selecting one representative Lie 2-algebroid. Even though the conditions that fixed the connection are geometrically meaningful, only the condition that it is $\eta$-compatible is a necessary one. For any other choice, an $\eta$- and $\mathcal{H}$-compatible generalised connection with a different torsion than the one we presented here exists such that the generalised gravity action for the new representative Lie 2-algebroid is the closed string effective action. Although we do not present a mathematical proof of this statement, it is supported by symbolic algebra calculations for generic choice of $\nabla$. In particular, there exist choices of $\nabla$ for which the generalised connection has vanishing Gualtieri tensor, whereas the natural unimodular one we chose here is not in this class.     

\section{Conclusions}

We have revisited the problem of defining torsion and Riemann curvature tensors in generalised geometry using a new approach. In  contrast to previous approaches, our starting point is not the usual data of a Courant algebroid $E$. In a number of previous approaches the Dorfman bracket, or its skew-symmetrized version, the Courant bracket, are  used to define geometric quantities that simulate the torsion and curvature tensors of ordinary differential geometry, with correction terms subsequently added in an ad hoc manner to restore linearity. In other approaches, the Ricci tensor is defined directly, without passing through a well-defined Riemann tensor. 

We  resolve these issues by harnessing the structure of Lie 2-algebroids. A Lie 2-algebroid contains by definition a binary bracket which is skew-symmetric and satisfies the Leibniz rule for functions in both entries; it is a generalised Lie bracket with the right properties to define bona fide torsion and curvature tensors. A Lie 2-algebroid can be constructed directly from Roytenberg's graded geometric description of Courant algebroids via symplectic dg-manifolds of degree 2. 
The construction relies on a split graded vector bundle structure, which corresponds geometrically to a choice of ordinary connection on~$E$. This choice should not be seen as a restriction; it is rather natural and justified on both mathematical and physical grounds: Mathematically, all Lie 2-algebroids obtained in this way are essentially equivalent (quasi-isomorphic). From the physics perspective, for dynamical problems such as gravity actions based on a generalised metric, such connections are natural and always available.

This construction brings us as close as possible to an analogue of the fundamental theorem of Riemannian geometry. For any choice of representative Lie 2-algebroid structure, there exists a unique generalised connection which is generalised metric compatible and has a fixed torsion, e.g. zero. In particular, we have presented a generalisation of the familiar Koszul formula of Riemannian geometry. A different representative, obtained by another choice of ordinary connection on $E$, gives a generalised connection with different torsion; this can be tracked through $B$-transforms. 

We have used this new point of view to write an Einstein-Hilbert-type action for Lie 2-algebroids. We have introduced a generalised metric and a volume form motivated by closed string theory (expressed as a section of the determinant bundle) and have constructed the generalised Ricci tensor for a generalised connection which is naturally defined via this metric. To simplify our analysis, we have considered a representative Lie 2-algebroid for the generalised tangent bundle by choosing a distinguished ordinary connection that satisfies sensible geometric conditions (compatibility with the inner and generalised metrics, and torsion along vector fields such that the unimodularity of the connection is respected.) We have found the unique generalised connection that corresponds to this Lie 2-algebroid for a specific choice of torsion that leads to an $O(D,D)$-covariant action. This procedure identifies the Lie 2-algebroid Einstein-Hilbert action with the low-energy effective action for closed strings, without imposing any restriction on the number of dimensions in this purely geometric setup.

This is an economical way of describing low-energy effective string actions as generalised geometry actions, avoiding the structural circularity and resulting ambiguities that affect previous work in the literature. It avoids undetermined components of the connection, leveraging a choice of a split dg-manifold, namely a graded vector bundle. From the viewpoint of physics, working in the category of (graded) vector bundles is a natural choice, even though it requires a choice of auxiliary geometric structure, which at the end of the day should not influence the physical results. At the same time, our approach confers two further advantages. First, the tight structure of a Lie 2-algebroid results directly in suitable generalisations of first (algebraic) and second (differential) Bianchi identities for the Riemann tensor. Second, the Riemann tensor is well-defined and one can consider higher-order curvature invariants, such as a proper definition of the Gauss-Bonnet term in generalised geometry. This is relevant for higher-derivative corrections to the closed string effective action. 

\begin{acknowledgments} 
Useful discussions with T.~Basile, P.~Batakidis,  M.~Cueca, F.~Hassler, N.~Ikeda, B.~Jur\v{c}o, J.~Kupka, E.~Lescano, J.~Moerland, F.~Petalidou, D.~Roytenberg, F.~Valach,  and J.~Vysok\'y are gratefully acknowledged. This work is funded by the Croatian Science Foundation project ``Higher Structures and Symmetries in Gauge and Gravity Theories'' (IP-2024-05-7921), by the European Union - NextGenerationEU, and by the Deutsche Forschungsgemeinschaft (DFG, German Research Foundation) -~545733782. This article is based upon work from the COST Action
21109 CaLISTA, supported by COST (European Cooperation in
Science and Technology). 
The work of Ath. Ch. was partially supported in its early stages by the Ulam Programme of the Polish National Agency for Academic Exchange.
The work of CH was supported by the STFC Consolidated Grant ST/X000575/1 and the STFC Consolidated Award UKRI4230.
We have profited from the symbolic computer algebra system Cadabra 2 in doing the computations for this paper \cite{Peeters:2018dyg}.
\end{acknowledgments}


\bibliography{amssamp}
\bibliographystyle{apsrev4-2}

\end{document}